\documentclass[reprint,superscriptaddress,amsmath,amssymb,aps,prl,longbibliography]{revtex4-2}

\usepackage{graphicx}
\usepackage{dcolumn}
\usepackage{bm}
\usepackage[pdftex,breaklinks,colorlinks=true,citecolor=blue,linkcolor=blue,urlcolor=blue]{hyperref}
\usepackage{ulem}
\usepackage{tabularx}
\usepackage{braket}
\usepackage{xcolor}
\usepackage{textcomp}

\begin{document}

\title{Assembly of a rovibrational ground state molecule in an optical tweezer}

\author{William~B.~Cairncross} 
\thanks{W.~B.~C.~and J.~T.~Z.~contributed equally to this work.}
\affiliation{Department of Physics, Harvard University, Cambridge, Massachusetts 02138, USA}
\affiliation{Department of Chemistry and Chemical Biology, Harvard University, Cambridge, Massachusetts 02138, USA}
\affiliation{Harvard-MIT Center for Ultracold Atoms, Cambridge, Massachusetts 02138, USA}

\author{Jessie~T.~Zhang} 
\thanks{W.~B.~C.~and J.~T.~Z.~contributed equally to this work.}
\affiliation{Department of Physics, Harvard University, Cambridge, Massachusetts 02138, USA}
\affiliation{Department of Chemistry and Chemical Biology, Harvard University, Cambridge, Massachusetts 02138, USA}
\affiliation{Harvard-MIT Center for Ultracold Atoms, Cambridge, Massachusetts 02138, USA}

\author{Lewis~R.~B.~Picard} 
\affiliation{Department of Physics, Harvard University, Cambridge, Massachusetts 02138, USA}
\affiliation{Department of Chemistry and Chemical Biology, Harvard University, Cambridge, Massachusetts 02138, USA}
\affiliation{Harvard-MIT Center for Ultracold Atoms, Cambridge, Massachusetts 02138, USA}

\author{Yichao~Yu} 
\affiliation{Department of Physics, Harvard University, Cambridge, Massachusetts 02138, USA}
\affiliation{Department of Chemistry and Chemical Biology, Harvard University, Cambridge, Massachusetts 02138, USA}
\affiliation{Harvard-MIT Center for Ultracold Atoms, Cambridge, Massachusetts 02138, USA}

\author{Kenneth~Wang} 
\affiliation{Department of Physics, Harvard University, Cambridge, Massachusetts 02138, USA}
\affiliation{Department of Chemistry and Chemical Biology, Harvard University, Cambridge, Massachusetts 02138, USA}
\affiliation{Harvard-MIT Center for Ultracold Atoms, Cambridge, Massachusetts 02138, USA}

\author{Kang-Kuen~Ni} 
\email{ni@chemistry.harvard.edu}
\affiliation{Department of Chemistry and Chemical Biology, Harvard University, Cambridge, Massachusetts 02138, USA}
\affiliation{Department of Physics, Harvard University, Cambridge, Massachusetts 02138, USA}
\affiliation{Harvard-MIT Center for Ultracold Atoms, Cambridge, Massachusetts 02138, USA}

\newcommand{\cSig}{c^3\Sigma_1}
\newcommand{\aSig}{a^3\Sigma_1}
\newcommand{\XSig}{X^1\Sigma}
\newcommand{\Na}{{\rm Na}}
\newcommand{\Cs}{{\rm Cs}}
\newcommand{\MHz}{{\rm MHz}}
\newcommand{\GHz}{{\rm GHz}}
\newcommand{\kHz}{{\rm kHz}}
\newcommand{\atomState}{\ket{3,3}_\Cs \ket{1,1}_\Na}
\newcommand{\Mtot}{M_{\rm tot}}
\newcommand{\mW}{{\rm mW}}
\newcommand{\um}{\text{\textmu}{\rm m}}
\newcommand{\us}{\text{{\textmu}s}}
\newcommand{\BPi}{B^1\Pi}
\newcommand{\bPi}{b^3\Pi}

\begin{abstract}
We demonstrate the  coherent creation of a single NaCs molecule in its rotational, vibrational, and electronic (rovibronic) ground state  in an optical tweezer. Starting with a weakly bound Feshbach molecule, we locate a two-photon transition via the $\ket{\cSig,v'=26}$ excited state and drive   coherent Rabi oscillations between the Feshbach state and a single  hyperfine level of the NaCs rovibronic ground state $\ket{\XSig,v''=0,N''=0}$ with a binding energy of $D_0 = h \times 147038.30(2)~\GHz$. We measure a lifetime of $3.4\pm1.6\,{\rm s}$ for the rovibronic ground-state molecule, which possesses a large molecule-frame dipole moment of 4.6~Debye and occupies predominantly the motional ground state. These long-lived, fully quantum-state--controlled individual dipolar molecules provide a key resource for molecule-based quantum simulation and information processing. 
\end{abstract}

\maketitle

Trapped arrays of individually controlled interacting atoms have enabled a range of  studies in quantum information and quantum many-body physics that are now reaching beyond what can be computed on classical machines~\cite{Bloch2012,Blatt2012, Zhang2017,Rispoli2019,Scholl2020,Ebadi2020}.  Substituting atoms with polar molecules, which feature rich internal states with tunable long-range dipolar interactions, further expands the opportunities for quantum simulation of novel phases of matter~\cite{Gorshkov2011b,levinsen2011topological,Baranov2012, Wall2015,Yao2018}, high fidelity quantum information processing~\cite{DeMille2002,Ni2018,Hudson2018, Albert2020}, precision measurements \cite{Kozyryev2017a,Kondov2019,Lim2018}, and studies of cold chemistry \cite{Ospelkaus2010,Klein2017,hu2019direct}. However, this molecular complexity also presents challenges in obtaining the same level of control in molecules as in atoms, motivating the development of new approaches~\cite{Danzl2008,Ni2008, Lang2008,Carr2009,Barry2014, Prehn2016,Kozyryev2017,Tarbutt2019,Valtolina2020}. Recent experiments have now realized coherent control of molecules at the level of individual particles and single internal quantum states for ions through cooling and readout of co-trapped atoms~\cite{Wolf2016, Chou2017} and with neutrals in optical tweezers~\cite{Liu2018,Anderegg2019,Zhang2020, Cheuk2020,He2020,Yu2020}. In select cases, motional control of single molecules has also been attained.

Associating single molecules from individual atoms cleanly maps the full quantum state control that is attainable for the constituent atoms onto the molecules. By controlling the motional states of the atoms prior to association, both the motional and the internal state of the resultant molecule can be controlled. Previously, full control has been demonstrated for single molecules in weakly bound states~\cite{Zhang2020,He2020,Yu2020}. In order to realize tunable dipolar interactions---the key ingredient for simulating strongly correlated, novel phases of matter---these single molecules must be transferred to a low-lying vibrational state where they possess an appreciable molecule-frame dipole moment.

\begin{figure}[tb]
    \centering
    \includegraphics[]{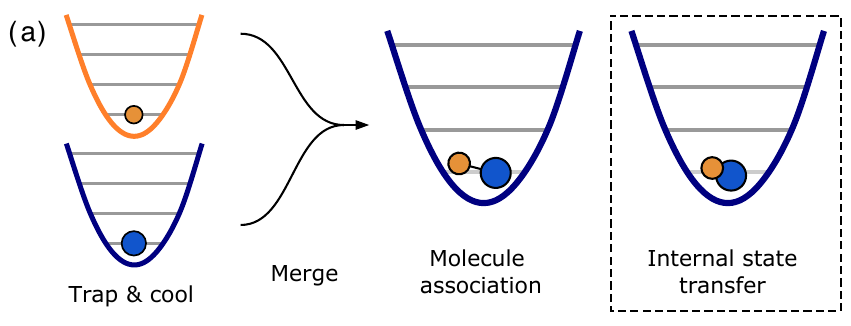}
    \includegraphics[]{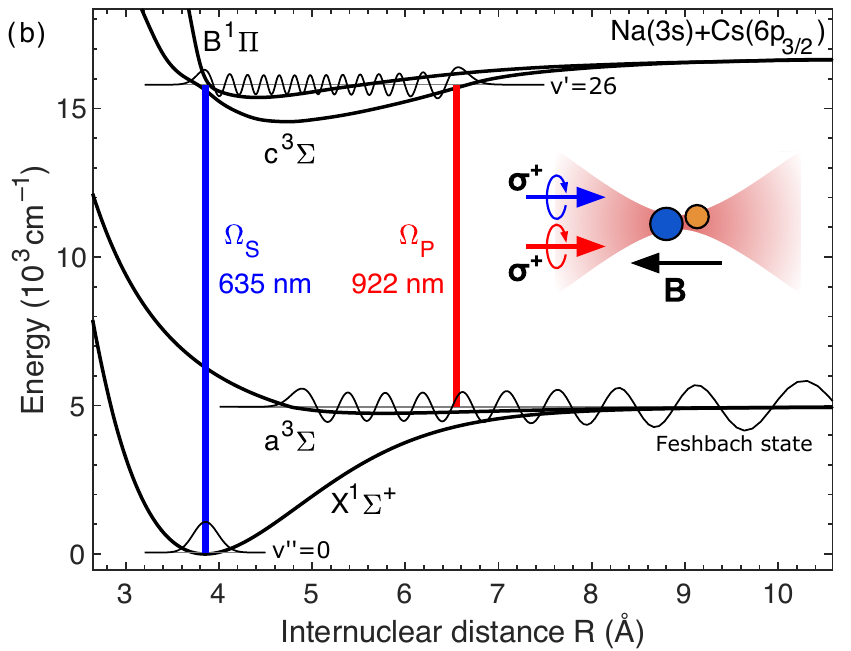}    
    \caption{Schematic overview of assembling rovibronic ground state molecules in optical tweezers. (a) Sequence of molecular assembly from individually trapped atoms to ground state molecule. This work focuses on the outlined final step of internal state transfer. (b) Selected potential curves of the NaCs molecule, showing the Raman transfer scheme from the Feshbach state to the rovibronic ground state via $\ket{\cSig,v'=26}$. Pump and Stokes laser Rabi frequencies are labeled $\Omega_P$ and $\Omega_S$, respectively. Inset: Geometry of the optical tweezer, magnetic bias field, and Raman transfer lasers. }
    \label{fig:fig1}
\end{figure}

In this Letter, we demonstrate the coherent association of a single rotational, vibrational, and electronic (rovibronic) ground-state molecule in an optical tweezer. Specifically, we perform ``molecular assembly'' [Fig.~\ref{fig:fig1}(a)] to form a single ground-state NaCs molecule starting from individually trapped single atoms that are first magnetoassociated into a weakly bound Feshbach molecular state. We transfer the Feshbach molecules to the rovibronic ground state using a coherent two-photon detuned Raman pulse, demonstrating internal state transfer of ultracold molecules in a new parameter regime. The resulting molecule possesses a large molecule-frame dipole moment of 4.6~Debye~\cite{Dagdigian1972,Dieglmayr2008}, and we measure a trapped  lifetime of $3.4 \pm 1.6$~s, which is limited by photon scattering. An inherent feature of the assembly process is that these molecules are created predominantly in the motional ground state of the optical tweezer. With the final step of ``molecular assembly'' completed, we establish a new platform for quantum simulation of novel phases of matter with arrays of molecules in the motional ground state of optical tweezers.

Our apparatus and experimental sequence have been described in detail previously \cite{Liu2018,Liu2019,Zhang2020}. First, single $^{23}$Na and $^{133}$Cs atoms are loaded from a dual-species magneto-optical trap into tweezers at 700~nm and 1064~nm, respectively [Fig.~\ref{fig:fig1}(a), first panel]. The atoms are loaded stochastically, and an initial non-destructive imaging step allows for postselection of the $\sim 30\%$ of experimental sequences where both atoms are initially present. Both atoms are then cooled to their respective 3D motional ground states using polarization gradient cooling followed by Raman sideband cooling. After cooling, we bring the atoms together into a single trap, and then perform Feshbach magnetoassociation to convert the atom pair into a weakly bound molecule \cite{Zhang2020} [Fig.~\ref{fig:fig1}(a), center panel]. The Feshbach molecule is formed in a single internal state with a binding energy of $\sim 1~\MHz$ at 863.7~G, and occupies predominantly the center-of-mass (COM) ground state of the optical tweezer. The experiment cycle time is typically $\sim 300~{\rm ms}$.

Next, we proceed to transfer the molecule from the Feshbach state to the rovibronic ground state $\ket{\XSig,v''=0,N''=0}$ [Fig.~\ref{fig:fig1}(a), last panel], where $v''$ and $N''$ are the vibrational and rotational quantum numbers, respectively, of $\XSig$. As in earlier work that used ensembles of molecules, we employ a two-photon optical transfer via an electronically excited state [Fig.~\ref{fig:fig1}(b)] \cite{Ni2008, Danzl2008, Lang2008, Chotia2012, Takekoshi2014, Molony2014, Park2015, Guo2016}. Earlier works have used stimulated Raman adiabatic passage (STIRAP) to transfer population from the Feshbach state. In STIRAP, modulation of pump and Stokes laser intensities as a function of time causes a coherent dark state to adiabatically evolve from the initial to final states while the intermediate state remains nearly unpopulated, thus minimizing scattering. In this work, we instead implement a detuned Raman transfer, and by doing so we demonstrate that high-fidelity creation of ground state molecules is possible in a different parameter regime than previously explored. The relevant parameters for a coherent transfer are the excited state linewidth $\Gamma$, the pump and Stokes Rabi frequencies $\Omega_{P,S}$, and the ratio $\Omega_R/R_{\rm sc}$, where $\Omega_R$ is the Raman Rabi frequency and $R_{\rm sc}$ is the (detuning-dependent) scattering rate. After locating the pump and Stokes transitions, we investigate these properties to find suitable parameters to perform Raman transfer.

\begin{figure*}[tb]
    \centering
    \includegraphics[width=0.75\textwidth]{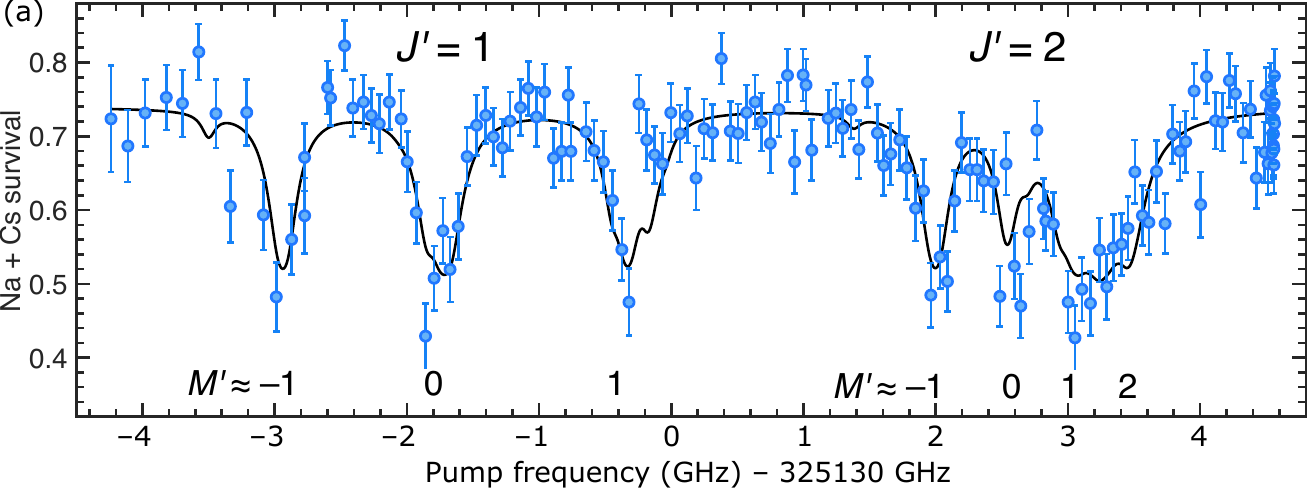} \\ \vspace{0.1in}
    \includegraphics[width=0.375\textwidth]{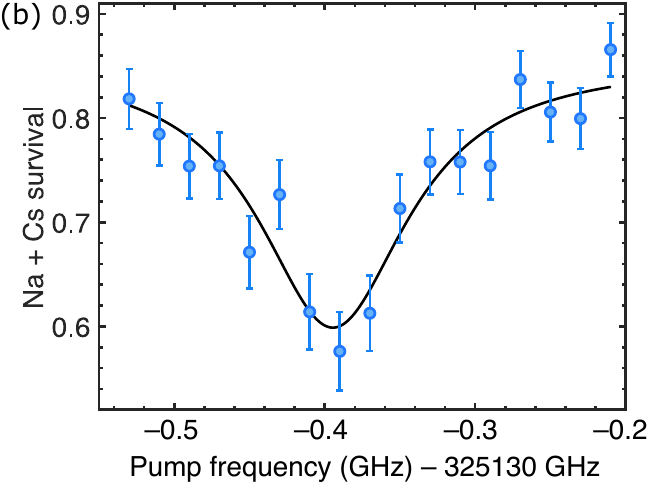}
    \includegraphics[width=0.368\textwidth]{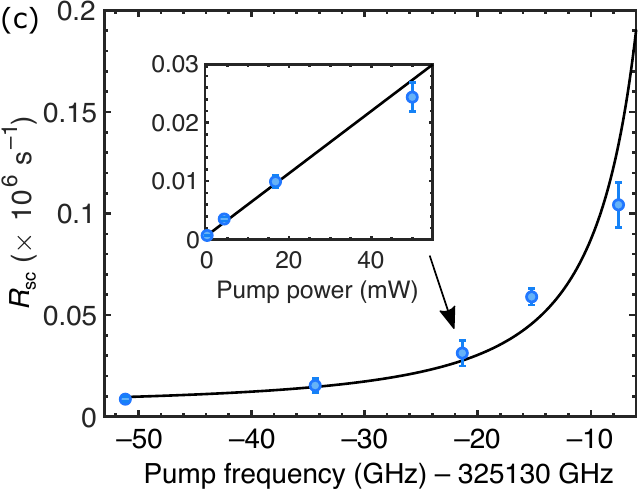}
    \caption{Characterization of the $\ket{\cSig,~v'=26}$ state of NaCs by depletion of Feshbach molecules. (a) Depletion spectrum in the vicinity of 325130~GHz using $\sigma^+ + \sigma^-$ polarization, showing the $J'=1,2$ rotational levels and Zeeman structure with approximate quantum number assignments. The curve is a theory prediction based on a model of the Feshbach and $\cSig$ states \cite{PRAtoBePublished}. (b) Linewidth characterization of the $J'=1,M'_J=1$ level of $\ket{\cSig,~v'=26}$ used for Raman transfer when probed with $\sigma^+$ polarization. The curve is a fit to a Lorentzian lineshape, with fit linewidth $\Gamma=2\pi \times 120(30)~\MHz$. (c) Characterization of the single-photon scattering rate $R_{\rm sc}$ due to the pump laser when red-detuned from $\ket{\cSig,~v'=26}$. The curve is a fit to the scattering rate derived from the same model as in part (a). Inset: the linear dependence of the scattering rate on the pump laser power at the $-21$~GHz detuning is consistent with single-photon scattering.}
    \label{fig:fig2}
\end{figure*}

A two-photon transfer to the ground state has not been previously performed in NaCs, necessitating first a search for and characterization of intermediate states as well as locating the ground state resonance. We choose $\ket{\cSig,v'=26}$ as an intermediate state due to several factors: (1)~it has relatively high Franck-Condon overlap with the Feshbach state, (2)~it is expected to have strong transition dipole moments to both the Feshbach and rovibronic ground states due to the large spin-orbit coupling constant of the Cs atom, and (3)~it is accessible from these states with convenient laser wavelengths of 922~nm and 635~nm, respectively [Fig.~\ref{fig:fig1}(b)]. Following a prediction based on the potential curves of Ref.~\cite{Grochola2011}, we located $\ket{\cSig,v'=26}$ using photoassociation spectroscopy. Figure \ref{fig:fig2}(a) shows a high-resolution spectrum of the transition from the Feshbach state to the $J'=1$ and $J'=2$ manifolds along with a model that incorporates the excited state structure, where $J'$ is the rotational plus electronic angular momentum in the $\cSig$ state. We then located the $\ket{\XSig,~v''=0}$ state using Autler-Townes spectroscopy. The details of the modeling and quantum number assignment are reported elsewhere~\cite{PRAtoBePublished}.

The linewidth of the intermediate state has a significant influence on our state transfer scheme. In earlier molecular association experiments, the molecular excited state used as an intermediate had a width comparable to that of the atomic transition to which it is asymptotically connected. In our system, that would be the Cs 6s $\rightarrow$ 6p transition, with a natural linewidth of order $\Gamma_{\rm atom}/2\pi \approx  5~\MHz$. However, we measure the linewidth for $\ket{\cSig,~v'=26,J'=1,M'_J=1}$ to be $\Gamma/2\pi\,=\,120(30)~\MHz$---more than an order of magnitude larger than the atomic linewidth [Fig.~\ref{fig:fig2}(b)], where $M'_J$ is the projection of $J'$ onto the laboratory magnetic field. We have not been able to determine the origin of the increased linewidth, and further investigation is warranted. We also characterized the scattering arising from all $\ket{\cSig,~v'=26}$ lines when red-detuned from resonance, as shown in Fig.~\ref{fig:fig2}(c). We find scattering rates consistent with our independent measurements of the $\ket{\cSig,v'=26}$ linewidth and transition strength, with the addition of a background scattering rate of $[200(100)~\us]^{-1}$ that may arise from further-detuned states.

While the observed linewidth of the $\cSig$ state is not ideal for state transfer, we find that the strength of the pump transition offsets this issue. We characterize the strength of the 922~nm pump transition by measuring the depletion time on the $J'=1,M'_J=1$ resonance at 325129.64(2)~GHz at low laser power, where the lifetime $\tau$ is long compared to $1/\Gamma$. In this limit, $\tau \approx \Gamma/\Omega_P^2$. We independently calibrate the pump laser intensity using a measurement of the vector AC Stark shift of the Cs hyperfine ground state, allowing us to extract the transition dipole moment. We find a Rabi frequency $\Omega_P/2\pi\,=\,6.2(8)~\MHz \times \sqrt{P_P/(1~\mW)}$ where $P_P$ is the pump laser power, corresponding to a transition dipole moment of $\mu_P=0.009(1)\,e\,a_0$. We attribute this large transition strength (when compared to other species of Feshbach molecules, {\it cf}.~\cite{Ni2008}) to the closed-channel dominated character of the Feshbach state. The tight confinement provided by the diffraction-limited optical tweezer allows us to use a small $\sim 13~\um$ waist for the Raman lasers, allowing Rabi frequencies up to $\sim 60~\MHz$ for this transition with $P_P\approx 100~{\rm mW}$.

Having characterized the excited state properties, we can then identify the specific $\cSig$ levels whose quantum numbers allow a two-photon transition to the rovibronic ground state. In the presence of a large magnetic bias field, the initial (Feshbach), intermediate ($\cSig$), and final ($\XSig$) states of our transfer scheme are in very different angular momentum coupling regimes. As a result, both initial and final states can couple to many excited states. Only a few intermediate levels couple to both and can thus give rise to Raman transitions, while others contribute only to loss via one-photon scattering. We find that choosing both lasers to have $\sigma^+$ polarization [Fig.~\ref{fig:fig1}(b) inset] gives the simplest spectrum within the current geometrical constraints of our apparatus. In this case, the Raman transition amplitude is dominated by the $\ket{J'=1,M'_J=1}$ state of $\ket{\cSig,~v'=26}$, connecting the Feshbach state to the $\ket{M''_{I_\Na}=3/2, M''_{I_\Cs}=5/2}$ hyperfine component of the $\ket{\XSig, N''=0}$ rotational state, where $M''_I$ is the projection of the nuclear spin $I$ onto the laboratory magnetic field. With the pump laser fixed on the $J'=1,M'_J=1$ resonance, we located the Stokes transition at 472166.04(2)~GHz using Autler-Townes spectroscopy. This measurement provides a value of the NaCs binding energy $D_0=147038.30(2)~\GHz$, improving on the previous precision by two orders of magnitude. 

\begin{figure*}[t]
    \centering
    \includegraphics[]{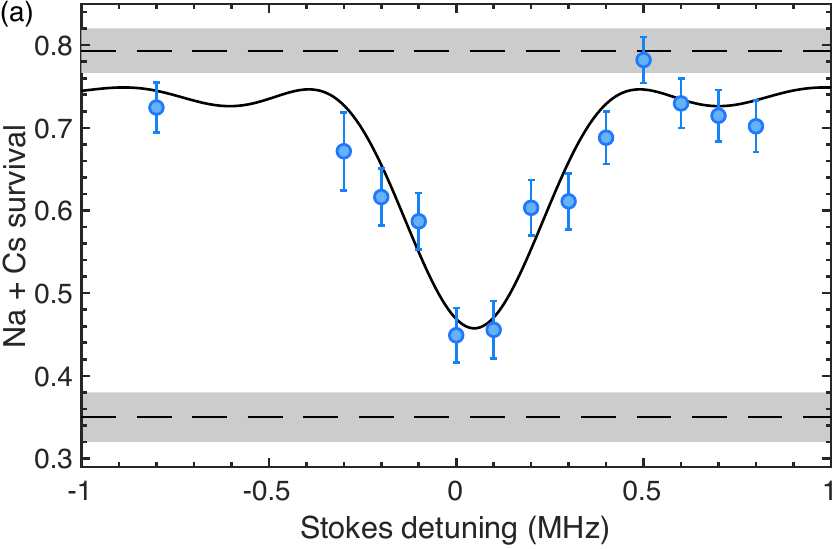} \hfill
    \includegraphics[]{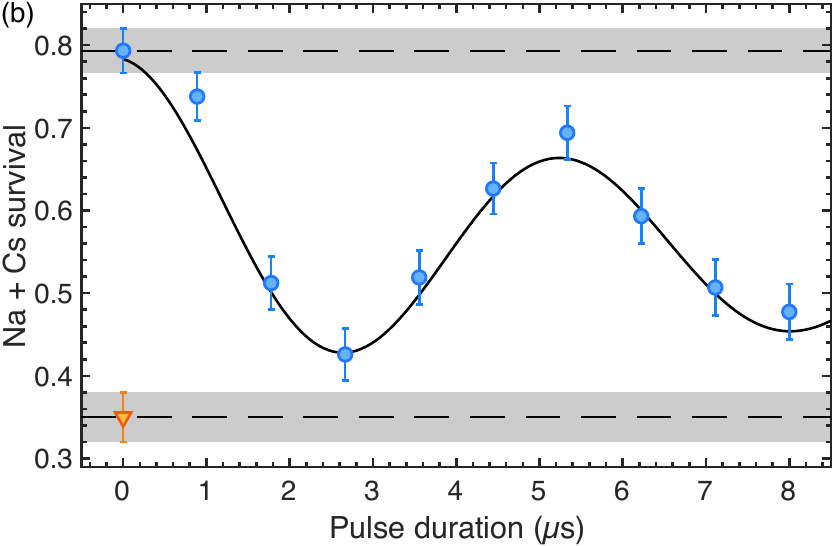}
    \caption{(a) Raman resonance and (b) Raman Rabi oscillation between Feshbach state and the rovibronic ground state of NaCs via $\ket{\cSig,v'=26}$. Dashed lines and gray bars show the Feshbach molecule contrast, which was collected simultaneously with the data in (b) and is shown there by the first blue circle and orange triangle. The data are simultaneously fit to a Rabi lineshape including loss and decoherence.}
    \label{fig:fig3}
\end{figure*}

Fixing the Stokes beam on resonance with $\ket{\cSig,J'=1,M'_J=1}$, we scan the pump beam frequency to calibrate the Stokes laser Rabi frequency $\Omega_S$ and the $\XSig \rightarrow \cSig$ transition dipole moment. As a function of pump detuning, we observe a characteristic Autler-Townes doublet, and we find $\Omega_S/2\pi\,=\,158(8)~\MHz \times \sqrt{P_S/(1~\mW)}$, where $P_S$ is the Stokes laser power \cite{supplement}. Using the estimated peak intensity of the Stokes beam, we obtain a transition dipole moment of $\mu_S\,=\,0.23(1)\,e\,a_0$. We find that the singlet fraction of the $\ket{\cSig, v'=26}$ state is $32(3)\%$; a substantial admixture which we attribute to the large spin-orbit coupling in the Cs 6p state and the close proximity of $\BPi$ vibrational states.

To perform Raman transfer, we detune both pump and Stokes lasers by $21~\GHz$ to the red of the $\ket{\cSig,v'=26}$ manifold. With a $3~\us$ pulse, we locate the Raman resonance shown in Fig.~\ref{fig:fig3}(a) by scanning the Stokes laser detuning. We observe only a single resonance, consistent with our expectation from calculations based on Ref.~\cite{Aldegunde2017} that we populate predominantly the $\ket{M''_{I_\Na}=3/2,M''_{I_\Cs}=5/2}$ hyperfine component of the rovibronic ground state for our choice of laser polarizations. Fixing the Stokes laser frequency on resonance and varying the pulse duration, we observe coherent Rabi oscillations [Fig.~\ref{fig:fig3}(b)]. We measure a Raman Rabi frequency of $\Omega_R/2\pi = 187(2)~\kHz$, consistent with the theoretical value of $210(30)~\kHz$ with pump Rabi frequency $\Omega_P/2\pi \approx 44(6)~\MHz$ at $P_P=50~\mW$, and Stokes Rabi frequency $\Omega_S/2\pi \approx 230(10)~\MHz$ at $P_S=2.1~\mW$. We find a ratio $\Omega_R/R_{\rm sc} = 27(7)$, indicating that coherent transfer dominates over loss.

We find a one-way transfer efficiency of 82(10)\% from Feshbach molecules to rovibronic ground state molecules. Incorporating the present Feshbach molecule creation fidelity of 38(1)\%, the overall efficiency for creation of ground-state molecules from individual atoms is $31(4)\%$, and the round-trip efficiency from atoms to ground state molecules and back is $25(4)\%$. The dominant factor limiting the overall molecule creation fidelity from atoms is that of Feshbach molecule creation, which is currently limited by heating of the atoms during the trap merge step. 

Figure~\ref{fig:fig3}(b) shows that another significant factor limiting ground-state molecule formation is decoherence. We fit a dephasing time of $\gamma^{-1} = 17(5)~\us$, while the scattering time is $R_{\rm sc}^{-1} = 23(6)~\us$. The observed decoherence can be accounted for by fluctuating AC Stark shifts arising from drifts in the pump and Stokes laser intensities. At present, the optical power in each of these beams drifts by $\sim 5\%$ due to thermal variation in the laboratory environment. For the data shown here, these powers were not actively stabilized. In planned improvements to our apparatus, we will actively stabilize the Raman laser powers so that the transfer efficiency will be limited by off-resonant scattering from $\cSig$. 

\begin{figure}[tb]
    \centering
    \includegraphics[]{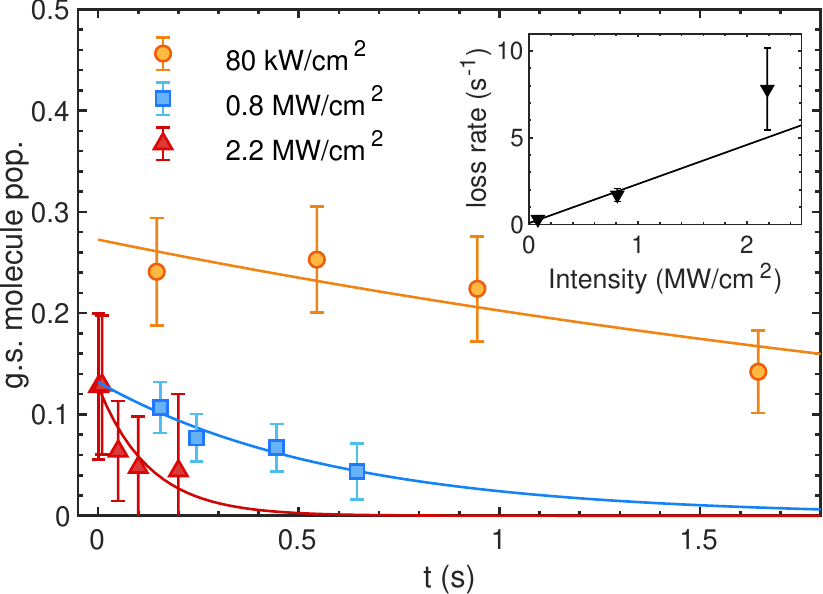}
    \caption{Characterization of the lifetime of a rovibronic ground state NaCs molecule in an optical tweezer at different values of the trap intensity. Atomic background has been subtracted. Inset: the increase in molecule loss rate with laser intensity is consistent with a linear trend corresponding to one-photon scattering.
    }
    \label{fig:fig4}
\end{figure}

Since our experiment involves only a single molecule in a deep optical trap, we expect the ground state lifetime to be primarily limited by scattering of the trap light or collisions with background gas. At our typical trap intensity of $80~{\rm kW/cm}^2$, we find a ground-state lifetime of $3.4 \pm 1.6$~s. In order to investigate possible limits to the molecule lifetime, we increased the tweezer intensity to 10$\times$ and 27$\times$ (Fig.~\ref{fig:fig4}). We find that the rovibronic ground state lifetime can be reduced to 0.5(1)~s and 130(40)~ms at trap intensities of 0.8~MW/cm$^2$ and 2.2~MW/cm$^2$, respectively, consistent with a linear scaling. 

Assuming a linear dependence on trap intensity, we find a loss rate for ground state molecules of $2.3(5)~{\rm s}^{-1}({\rm MW/cm}^2)^{-1}$~(Fig.~\ref{fig:fig4} inset). The precision of our measurement is limited by a maximum cycle time of 1.5~s due to thermal fluctuations of the apparatus. Using the theoretical ground state polarizability of NaCs from Ref.~\cite{Vexiau2017}, we expect a scattering rate of $54~{\rm s}^{-1}({\rm MW/cm}^2)^{-1}$, suggesting either an overestimate of the theoretical polarizability or a high proportion of Rayleigh over Raman scattering at this wavelength.

The rovibronic ground state molecule primarily inherits the motional quantum state of the Feshbach molecule, which arises from the individually laser cooled atoms. We estimate that the rovibronic ground state molecule occupies the motional ground state with 65(5)\% probability, compared to 75(5)\% for the Feshbach molecule \cite{supplement}. This excitation of the motion of the rovibronic ground state molecule arises from two sources: First, the absorption of a pump photon and emission of a Stokes photon imparts a coherent momentum kick to the molecule, removing it from the ground state with $\sim 8\%$ probability. Second, the Feshbach and rovibronic ground states experience different optical trapping frequencies due to their differential polarizability, leading to a wavefunction mismatch that projects population onto excited motional states with $\sim 6\%$ probability. These effects can be mitigated by performing the Raman transfer more slowly with a larger detuning and at a higher trap frequency. However, loss of Feshbach molecules and finite laser coherence will limit the maximum transfer time. In the limit of resolved motional sidebands during Raman transfer, both effects can be eliminated. It is also possible to apply the Raman lasers perpendicular to the tweezer axis in order to take advantage of the higher trap frequencies in the radial direction. 

In summary, we have demonstrated the coherent creation of a single rovibronic ground state molecule predominantly in the motional ground state of an optical tweezer. The $31(4)\%$ overall fidelity of ground-state molecule production demonstrated here is not fundamentally limited at any step, and can be improved with optimization of the atom ground-state cooling and merging. 

This work completes the final step of molecular assembly in our work towards arrays of molecules in the motional ground state of optical tweezers. With a single quantum-state-controlled rovibrational ground state molecule in hand, recent demonstrations of scaling to larger arrays with atoms \cite{Scholl2020,Ebadi2020} then serve as a starting point for parallel molecular assembly in the near future, providing a platform for engineering controlled long-range entangling interactions between molecules. With the addition of established techniques for microwave and electric field control~\cite{Matsuda2020,Yan2020}, molecular qubits for quantum computing applications \cite{Ni2018,Albert2020} and simulations that further our understanding of quantum phases of matter \cite{Gorshkov2011b,levinsen2011topological,Baranov2012,Yan2013, Wall2015,Yao2018} are all within experimental reach.

\begin{acknowledgments}
We thank Lee R.~Liu and Constantin Arnscheidt for early experimental assistance.
We thank Jeremy Hutson, Rosario Gonz{\'a}lez-F{\'e}rez, Olivier Dulieu, and Eberhard Tiemann for helpful discussions and Robert Moszynski for providing {\it ab-initio} transition dipole moments of NaCs. This work is supported by the NSF~(PHY-1806595), the AFOSR~(FA9550-19-1-0089), and the Camille and Henry Dreyfus Foundation~(TC-18-003). J.~T.~Z. is supported by a National Defense Science and Engineering Graduate Fellowship. W.~B.~C. is supported by a Max Planck-Harvard Research Center for Quantum Optics fellowship. K.~W. is supported by an NSF GRFP fellowship. 
\end{acknowledgments}

\bibliography{master_ref}
\bibliographystyle{apsrev4-2}

\end{document}

\title{Supplemental Material: \\ Assembly of a rovibrational ground state molecule in an optical tweezer}

\author{William~B.~Cairncross} 
\thanks{W.~B.~C.~and J.~T.~Z.~contributed equally to this work.}
\affiliation{Department of Physics, Harvard University, Cambridge, Massachusetts 02138, USA}
\affiliation{Department of Chemistry and Chemical Biology, Harvard University, Cambridge, Massachusetts 02138, USA}
\affiliation{Harvard-MIT Center for Ultracold Atoms, Cambridge, Massachusetts 02138, USA}

\author{Jessie~T.~Zhang} 
\thanks{W.~B.~C.~and J.~T.~Z.~contributed equally to this work.}
\affiliation{Department of Physics, Harvard University, Cambridge, Massachusetts 02138, USA}
\affiliation{Department of Chemistry and Chemical Biology, Harvard University, Cambridge, Massachusetts 02138, USA}
\affiliation{Harvard-MIT Center for Ultracold Atoms, Cambridge, Massachusetts 02138, USA}

\author{Lewis~R.~B.~Picard} 
\affiliation{Department of Physics, Harvard University, Cambridge, Massachusetts 02138, USA}
\affiliation{Department of Chemistry and Chemical Biology, Harvard University, Cambridge, Massachusetts 02138, USA}
\affiliation{Harvard-MIT Center for Ultracold Atoms, Cambridge, Massachusetts 02138, USA}

\author{Yichao~Yu} 
\affiliation{Department of Physics, Harvard University, Cambridge, Massachusetts 02138, USA}
\affiliation{Department of Chemistry and Chemical Biology, Harvard University, Cambridge, Massachusetts 02138, USA}
\affiliation{Harvard-MIT Center for Ultracold Atoms, Cambridge, Massachusetts 02138, USA}

\author{Kenneth~Wang} 
\affiliation{Department of Physics, Harvard University, Cambridge, Massachusetts 02138, USA}
\affiliation{Department of Chemistry and Chemical Biology, Harvard University, Cambridge, Massachusetts 02138, USA}
\affiliation{Harvard-MIT Center for Ultracold Atoms, Cambridge, Massachusetts 02138, USA}

\author{Kang-Kuen~Ni} 
\email{ni@chemistry.harvard.edu}
\affiliation{Department of Chemistry and Chemical Biology, Harvard University, Cambridge, Massachusetts 02138, USA}
\affiliation{Department of Physics, Harvard University, Cambridge, Massachusetts 02138, USA}
\affiliation{Harvard-MIT Center for Ultracold Atoms, Cambridge, Massachusetts 02138, USA}

\maketitle

\onecolumngrid

% Add S1 to bibliography
\bibliographystyle{apsrev4-2}
\renewcommand*{\citenumfont}[1]{S#1}
\renewcommand*{\bibnumfmt}[1]{[S#1]}

%Add S to labels
\setcounter{table}{0}
\renewcommand{\thetable}{S\arabic{table}}%
\setcounter{figure}{0}
\renewcommand{\thefigure}{S\arabic{figure}}%
\renewcommand{\thepage}{S\arabic{page}}  
\renewcommand{\thesection}{S\arabic{section}}   
\renewcommand{\theequation}{S.\arabic{equation}}

\newcommand{\na}{\text{Na}}
\newcommand{\cs}{\text{Cs}}
\newcommand\com{\text{com}}
\newcommand\rel{\text{rel}}
\newcommand\acom{a_\text{com}}
\newcommand\arel{a_\text{rel}}
\newcommand\nbar{\bar{n}}

\newcommand{\cSig}{c^3\Sigma_1}
\newcommand{\XSig}{X^1\Sigma}
\newcommand{\aSig}{a^3\Sigma}
\newcommand{\Na}{{\rm Na}}
\newcommand{\Cs}{{\rm Cs}}
\newcommand{\MHz}{{\rm MHz}}
\newcommand{\GHz}{{\rm GHz}}
\newcommand{\kHz}{{\rm kHz}}
\newcommand{\atomState}{\ket{3,3}_\Cs \ket{1,1}_\Na}
\newcommand{\Mtot}{M_{\rm tot}}
\newcommand{\mW}{{\rm mW}}
\newcommand{\um}{\mu{\rm m}}
\newcommand{\us}{\mu{\rm s}}
\newcommand{\BPi}{B^1\Pi}
\newcommand{\bPi}{b^3\Pi}

\section{Pump and Stokes Rabi frequency calibration}

\begin{figure}[htb]
    \centering
    \includegraphics[]{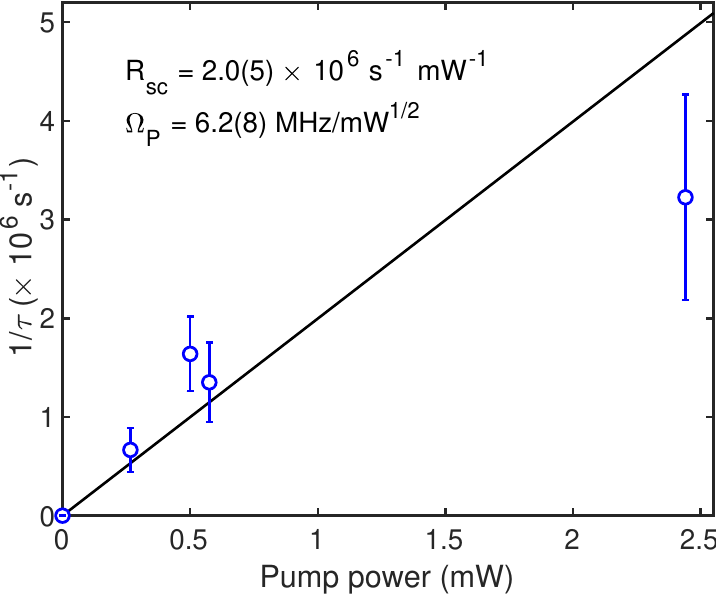}
    \caption{Depletion rate of Feshbach molecules as a function of pump laser power, for calibration of pump laser Rabi frequency.}
    \label{fig:pumpRabi}
\end{figure}

We calibrate the pump Rabi frequency by measuring the Feshbach (FB) molecules loss rate as a function of laser power when the laser is resonant with the $\ket{J'=1,M'_J=1}$ state. In the regime where $\Omega_P \ll \Gamma$, the depletion rate is $1/\tau \approx \Omega_P^2/\Gamma$. We fit the inverse FB molecule lifetime versus pump power as shown in Fig.~\ref{fig:pumpRabi}, and extract a Rabi frequency of $\Omega_P/2\pi = 6.2(8)~\MHz \times \sqrt{\frac{P_P}{1~\mW}}$.

\begin{figure}[htb]
    \centering
    \includegraphics[]{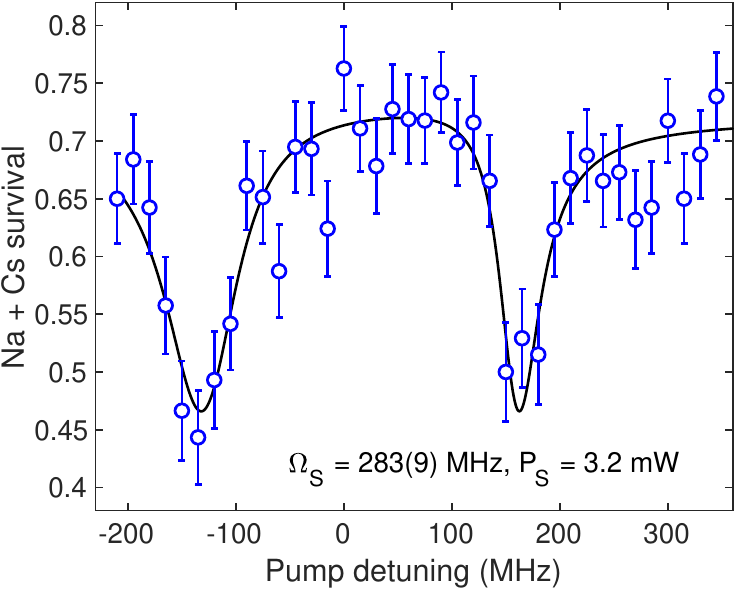}
    \caption{Autler-Townes splitting of $\ket{\cSig,v'=26,J'=1,M'_J=1}$ used for calibration of Stokes laser Rabi frequency.}
    \label{fig:ATdoublet}
\end{figure}

We calibrate the Stokes laser Rabi frequency using Autler-Townes spectroscopy. After locating the rovibrational ground state, we tune the Stokes laser onto resonance with the $\ket{J'=1,M'=1}$ state, and measure the depletion of Feshbach molecules as a function of pump laser frequency. We observe an Autler-Townes doublet feature as shown in Fig.~\ref{fig:ATdoublet}, and fit it to a model function for the 2-body survival probability $P_{\rm Na+Cs}$ after a pulse duration $t$ \cite{Fleischhauer2005}
\begin{equation}
    P_{\rm Na+Cs}(t) = P_{\rm atom} + P_{\rm mol} \exp{\left( - \Omega_P^2 t \frac{4 \Gamma (\Delta_P-\Delta_S)}{|\Omega_S^2 + 2 i (\Delta_P-\Delta_S) (\Gamma + 2 i \Delta_P)|^2} \right)},
\end{equation}
where $P_{\rm atom}$ is an atomic background population, $P_{\rm mol}$ is the Feshbach molecule creation fidelity, and $\Delta_{P,S}$ are the one-photon detunings of pump and Stokes lasers, respectively. We obtain a value of $\Omega_S/2\pi=283(9)~\MHz$ at $P_S=3.2~\mW$, or $\Omega_S/2\pi=158(8)~\MHz \times \sqrt{P_S/(1~\mW)}$ for the Stokes laser Rabi frequency. 

\section{Motional ground state fraction}
In Ref.~\cite{Zhang2020}, we estimated that the center of mass (COM) ground state fraction of Feshbach molecules was 77(5)\% using Raman sideband thermometry. In that work, we also observed a Feshbach molecule creation fidelity of 47(1)\%, consistent with a relative motional ground state population of 58(4)\%. Presently, we observe a slightly reduced Feshbach molecule creation fidelity of 38(3)\%, while the atomic ground state cooling conditions have not changed from Ref.~\cite{Zhang2020}. The reduced fidelity of Feshbach molecule creation arises from an axial misalignment of the dual species optical tweezers, which primarily leads to heating of the Na atom during the merge process. Because Na is much lighter than Cs, heating of the Na atom primarily contributes to excitation of the relative motional degree of freedom, as opposed to the COM. Using the Feshbach molecule creation fidelity as a proxy for thermometry, we estimate that the COM ground state fraction of Feshbach molecules under present conditions is 75(5)\%. In the following, we neglect state changes of the $\sim 25\%$ initially motionally excited molecules.

Two main effects cause COM motional excitation of the molecule during Raman transfer. The first arises from the differential wavenumber of the pump and Stokes lasers, $\Delta k$. In the Raman transfer process, motional sidebands of the COM degree of freedom are not resolved: The Raman Rabi frequency is of order $\Omega_R\approx 2\pi \times 200~\kHz$, while the axial trap frequency during transfer is $\omega_z\approx 2 \pi \times 3~\kHz$. Coherent absorption of a pump photon and emission of a Stokes photon during Raman transfer thus creates a coherent state of the molecule's COM motion with momentum $\hbar \Delta k$. The mean occupation number resulting from this kick is $\bar{n} \approx 0.085$, giving a 92\% probability to remain in the ground motional state.

A second source of motional excitation during Raman transfer is the mismatch of initial and final state trap frequencies, due to a difference in polarizability of each state to the trapping laser. The Feshbach state polarizability is estimated as ${\rm Re}[\alpha_{\rm Na+Cs}(1064~{\rm nm})] = 1397~a_0^3$ from the sum of atomic polarizabilities, while the $\XSig$ polarizability is ${\rm Re}[\alpha_{\XSig}(1064~{\rm nm})] = 936~a_0^3$ from \cite{Vexiau2017}. The ratio of trap depths is then $1397/936 \approx 1.5$, so that ratio of trap frequencies is $\sqrt{1.5}$ which we call $f$ below, and the ratio of harmonic oscillator lengths is $(1.5)^{1/4}$. The probability of remaining in the motional ground state is
\begin{equation}
    P_{0 \leftarrow 0} = |\bra{n'_x=0,n'_y=0,n'_z=0} e^{i \Delta k z} \ket{n_x=0,n_y=0,n_z=0}|^2.
\end{equation}
The initial Feshbach molecule COM spatial wavefunction is essentially projected onto the rovibrational ground state COM wavefunction. Because the Raman transfer lasers propagate along the $z$ axis, we can separate the integrals and evaluate them straightforwardly, finding
\begin{equation}
    P_{0\leftarrow 0} = \frac{8 f^{3/2}}{(f+1)^3} \exp{\left(- \frac{f \, \eta^2}{2f + 2}\right)},
\end{equation}
where $\eta = \Delta k \, z_{\rm ho} \approx 0.413$ is the Lamb-Dicke parameter. We find $P_{0\leftarrow 0} \approx 0.94$. 

In the worst case, in which we neglect any cancellation of these two motional excitation effects, we would expect a reduction of the motional ground state population by a factor 0.862. We estimate the overall COM ground state fraction of the rovibrational ground state to be 75(5)\% $\times$ 0.862 = 65(5)\%. Both excitation effects could be reduced by increasing the trap frequency in order to perform Raman transfer in a sideband-resolved regime, however this will be limited by scattering of both trap and Raman lasers. 

\bibliography{master_ref}